\documentclass[doublecol,linenumbers]{epl2} % for 2 columns style with line numbers
\usepackage{amssymb}
\usepackage{amsmath}
\usepackage{bm}

\title{Magnetization pumping and dynamics in a Dzyaloshinskii-Moriya
magnet}
\shorttitle{Magnetization pumping and dynamics in a Dzyaloshinskii-Moriya
magnet} %Insert here a short version of the title if it exceeds 70 characters

\author{Alexey A. Kovalev\inst{1} \and Utkan G\"{u}ng\"{o}rd\"{u}\inst{1}}
\shortauthor{Alexey A. Kovalev \etal}

\institute{                    
  \inst{1} Department of Physics and Astronomy and Nebraska Center for Materials
and Nanoscience, University of Nebraska, Lincoln, Nebraska 68588,
USA}
\abstract{
We formulate a phenomenological description of thin ferromagnetic
layers with inversion asymmetry where the single-domain magnetic
dynamics experiences magnon current-induced torques and leads to magnon-motive
forces. We first construct a phenomenological theory based on
irreversible thermodynamics, taking into account the symmetries of the system. Furthermore, we confirm that these effects originate from Dzyaloshinskii\textendash Moriya
interactions from the analysis based on the
stochastic Landau-Lifshitz-Gilbert equation. Our phenomenological results generalize to a general form of Dzyaloshinskii\textendash Moriya
interactions and to other systems, such as pyrochlore crystals and chiral magnets. Possible applications
include spin current generation, magnetization reversal and magnonic
cooling.}
\nolinenumbers

\begin{document}

\maketitle

\section{Introduction} 

Spincaloritronics studies various thermal effects
relying on the spin degree of freedom \cite{Goennenwein:mar2012}.
The most prominent examples are the spin Seebeck effect \cite{Uchida.Takahashi.ea:2008},
the spin Peltier effect \cite{Slachter.Bakker.ea:NP2010,Dubi.DiVentra:PRB2009},
and thermally induced motion of domain walls \cite{Hinzke:PRL2011,Kovalev:EPL2012}.
Apart from fascinating physics, in some instances related to appearance of spin-motive force\cite{Volovik:JPC1987}, these studies might offer new ways
for energy harvesting, cooling, and magnetization control \cite{Hatami.Bauer.ea:PRL2007,Kovalev:SSC2010}.
Spincaloritronics might also help with the development of electronics
relying on \textit{pure} spin currents \cite{Kovalev:PRB2009,Slachter.Bakker.ea:NP2010}
which contrasts conventional electronics.
The known ways to create pure spin currents include non-local spin
injection in spin valves,
optical injection by circularly polarized light,
spin pumping, and spin Hall effect\cite{Johnson.Silsbee:PRL1985}.

Pure spin currents in a form of magnon flow have attracted considerable
attention recently as they can transfer signals \cite{Kajiwara:mar2010}
and even realize magnonic logic circuits with low dissipation and
without generation of Oersted fields \cite{Khitun.Wang:JoAP2011}.
On the other hand magnons can exhibit similar phenomena to electrons,
e.g., spin-transfer torque on magnetic textures such as domain walls
\cite{Hinzke:PRL2011} and skyrmions
\cite{Kong:PRL2013,Kovalev:PRB2014}, Hall
effect, and topologically
protected edge states\cite{Onose:2010}.
Such magnonic spin currents can be driven by radio-frequency fields
or temperature gradients. In the ferrimagnetic insulator yttrium iron
garnet (YIG) magnons can travel over large distances without interruption
due to remarkably low Gilbert damping \cite{Douglass:PR1963}.

Relativistic effects result in interesting physics
in the context of spin currents. Discovery of spin-orbit torques
\cite{Fang.Kurebayashi.ea:NN2011} allows for magnetization control
by charge currents in bilayers consisting of a layer with strong spin-orbit
interactions, e.g., metal or topological insulator, and a ferromagnet
\cite{Miron:Nature2010}.
Spin-orbit torques are often interpreted in terms of a Rashba contribution
\cite{Miron:Nature2010} and a spin Hall contribution \cite{Liu:PRL2011}
while in a general scenario it is more useful to separate reactive
and dissipative contributions \cite{Tserkovnyak.Bender:PRB2014}.
Magnons on the other hand can be influenced by Dzyaloshinskii-Moriya
interactions (DMI) similar to how electrons are influenced
by spin-orbit interactions\cite{Costa.Muniz.ea:PRB2010,Moon.Seo.ea:PRB2013,Manchon.Ndiaye.ea:2014}.
In particular, a spin-orbit-like torque generated by spin waves has been suggested \cite{Manchon.Ndiaye.ea:2014}.
DMI can result from spin-orbit interactions in systems with broken
inversion symmetry \cite{Dzyaloshinsky:JoPaCoS1958}
or from structural asymmetry in ultrathin magnetic bilayers  \cite{Crepieux.Lacroix:JoMaMM1998}.
Thus, one can naturally expect magnon analogs of spin-orbit
torques and charge pumping observed in metal/ferromagnet bilayers
(see the figure). 

In this paper, we develop a general phenomenological description of
the interplay between magnetization dynamics, magnon currents, and temperature
gradients in single-domain ferromagnetic layers lacking inversion symmetry. We accompany
our analysis by a model based on the stochastic Landau-Lifshitz-Gilbert
(LLG) equation with DMI characteristic to ultrathin magnetic layers
with structural asymmetry. We obtain reactive and dissipative torques
on uniform magnetization and discuss the possibility to reverse magnetization
by magnon currents and temperature gradients. We discuss pumping of
magnonic spin currents by precessing single-domain magnetization and
analyze the feasibility of magnonic cooling.
\begin{figure}
\centerline{\includegraphics[width=0.8\columnwidth]{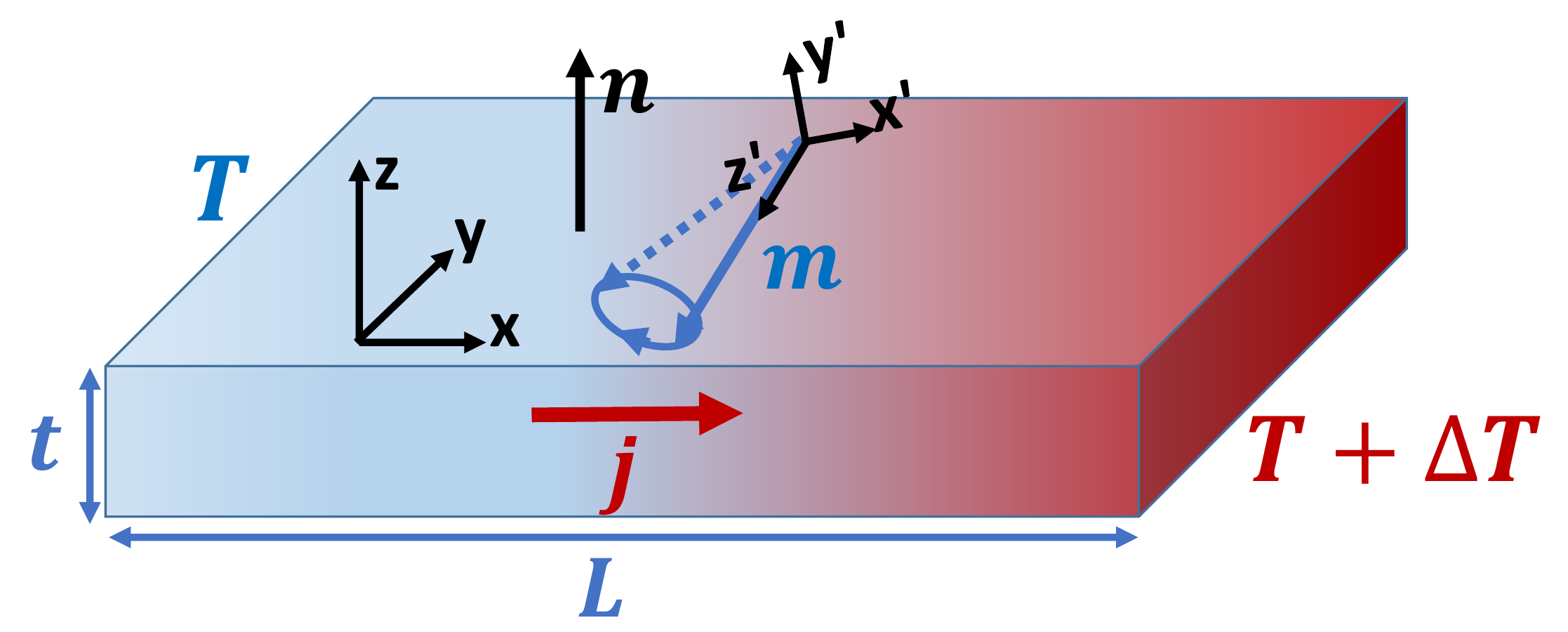}}
\protect\caption{(Color online) Single-domain magnetization dynamics induced by microwave
field pumps magnon $j$ and spin $j_{s}=-\hbar j$ currents by virtue
of Dzyaloshinskii-Moriya interactions. This can develop a temperature
gradient along the sample. Alternatively, a temperature gradient can
result in magnon current and torque on uniform magnetization according
to the Onsager reciprocity principle. }
\label{fig:Fig1} 
\end{figure}

\section{Phenomenology of thermal magnons with DMI} 

In this section, we employ general principles of non-equilibrium thermodynamics \cite{Landau.Lifshits:vol5} in order to formulate phenomenology of thermal magnons applicable to a system with interfacial inversion asymmetry as in the figure.  We begin by constructing a general
phenomenological description of magnonic and thermal currents in a single-domain
ferromagnet in which thermodynamic variables represent the
direction of the reduced (averaged over the magnonic excitations)
spin density $\mathbf{m}_{s}$ (for convenience the index $s$ is dropped in this
section), density of magnons $\rho$ and density of energy $\rho_{u}$.
We assume that the single-domain ferromagnet is taken out of equilibrium by applying
temperature and chemical potential gradients (non-zero chemical potential
for magnons can be created by, e.g., microwave pumping \cite{Demokritov:Nature2006}).
An appropriate equation of motion then determines how the ferromagnet evolves back towards equilibrium. We now
write the rate of the entropy production \cite{Landau:1984}: 
\begin{equation}
\dot{\mathbb{S}}=-\int d^{3}\mathbf{r}\,\frac{\boldsymbol{\partial}\cdot\mathbf{j}_{u}+\mu\dot{\rho}+\mathbf{H}_{\mathrm{eff}}\cdot\mathbf{\dot{m}}}{T}\:,\label{Entropy_Rate}
\end{equation}
where we introduced the magnon ($\mathbf{j}$) and energy ($\mathbf{j}_{u}$)
currents, and the conjugate/force corresponding to the spin density direction
defined as $-\delta_{\mathbf{m}}\mathbb{S}|_{\mathbf{j}_{U}(\mathbf{j})=0}=\mathbf{H}_\mathrm{eff}/T$.
It is convenient to introduce the modified
energy current $\mathbf{j}_{q}=\mathbf{j}_{u}-\mu\mathbf{j}$ in order to
arrive at the more familiar equation for the rate of the entropy production \cite{Landau:1984}:
\begin{equation}
\dot{\mathbb{S}}=\int d^{3}\mathbf{r}\left(-\frac{\boldsymbol{\partial} T}{T^{2}}\cdot\mathbf{j}_{q}-\frac{\boldsymbol{\partial}\mu}{T}\cdot\mathbf{j}-\frac{\mathbf{H}_\mathrm{eff}}{T}\cdot\mathbf{\dot{m}}\right).\label{Entropy_Rate_1}
\end{equation}
Here we integrated the term involving $\mathbf{j}_{q}$ by parts, used the local conservation laws of energy and
number of magnons, $\dot{\rho}=-\boldsymbol{\partial}\cdot\mathbf{j}-\rho/\tau$
and $\dot{\rho}_{u}=-\boldsymbol{\partial}\cdot\mathbf{j}_{u}$,
and disregarded the term $\mu\rho/\tau$ where $\tau$ corresponds to the life time of magnons.
This is possible when the number of magnons is approximately conserved.
One can also consider the opposite limit in which magnons quickly relax to the local equilibrium without the build up of large $\mu$ ($\mu\approx0$) as in this case the term $\mu\rho/\tau$ can be also disregarded. The remaining conjugates/forces
can be immediately identified as $-\delta_{\mathbf{j}_{q}}\dot{\mathbb{S}}|_{\mathbf{m},\mathbf{j}=0}=-\boldsymbol{\partial}\left(1/T\right)$
and $-\delta_{\mathbf{j}}\dot{\mathbb{S}}|_{\mathbf{m},\mathbf{j}_{q}=0}=\boldsymbol{\partial}\mu/T$. 

We now relate the currents $\mathbf{j}$ and $\mathbf{j}_{q}$ and the time derivative of the spin density direction
$\mathbf{\dot{m}}$ to the thermodynamic conjugates via kinetic coefficients.
By accounting for the structural asymmetries defined by the $\mathbf{n}$-axis
in the figure, we obtain the magnon/energy current expansion
in terms of the chemical potential and temperature gradients as well
as the magnetization dynamics responsible for fictitious fields on
the magnons: 
\begin{equation}
\begin{array}{c}
-\partial_\alpha\mu=\hat{\varrho}\mathbf{j}+\hat{\Pi}\boldsymbol{\partial}T/T-\left(\eta\mathbf{m}\times\mathcal{D}_\alpha\mathbf{m}+\vartheta\mathcal{D}_\alpha\mathbf{m}\right)\cdot\mathbf{\dot{m}}\,,\\
\\
(j_q)_\alpha=\hat{\Pi}^{T}\mathbf{j}-\hat{\kappa}\boldsymbol{\partial}T-\left(\eta_1\mathbf{m}\times\mathcal{D}_\alpha\mathbf{m}+\vartheta_1\mathcal{D}_\alpha\mathbf{m}\right)\cdot\mathbf{\dot{m}},
\end{array}\label{ohmD-1}
\end{equation}
where $\mathcal{D}_\alpha=(D/A)(\boldsymbol{n}\times\boldsymbol{e}_\alpha) \times$ is a part of the chiral derivative accounting for DMI, $\boldsymbol{e}_\alpha$ are basis vectors\cite{Tserkovnyak.Bender:PRB2014,Kim.Lee.ea:PRL2013}, $\eta$ 
and $\vartheta$ are the so-called reactive (also referred to as spin-motive force) and dissipative coefficients\cite{Tserkovnyak.Bender:PRB2014}
(generally $\mathbf{m}\cdot\mathbf{n}$ dependent), $\hat{\varrho}$,
$\hat{\Pi}$ and $\hat{\kappa}$ are the resistivity, Peltier and
thermal conductivity tensors, respectively, which are in general temperature
dependent. As it will become clear from the following discussion, it is convenient to invert the equation for the magnon current, as it has been done in Eq.~(\ref{ohmD-1}).  In the remaining part of the paper we will not consider
corrections corresponding to $\eta_{1}$ and $\vartheta_{1}$ as these
contributions do not appear in our microscopic
treatment. The axial symmetry around the $\mathbf{n}$-axis
leads to the separation of the conductivity tensor $\hat{g}\equiv\hat{\varrho}^{-1}$
and the thermal conductivity tensor $\hat{\kappa}$ into the longitudinal
$g$ and $\kappa$, and the Hall $g_{H}$ and $\kappa_{H}$ contributions
where $\hat{g}=g+g_{H}\mathbf{n}\times$ and $\hat{\kappa}=\kappa+\kappa_{H}\mathbf{n}\times$.
The LLG equation becomes:
\begin{equation}
\begin{array}{c}
\mathfrak{s}(1+\alpha\mathbf{m}\times)\mathbf{\dot{m}}+\mathbf{m}\times\mathbf{H}_\mathrm{eff}=\left(\eta+\vartheta\mathbf{m}\times\right)j_\alpha\mathcal{D}_\alpha\mathbf{m}\\
\\
+\left(\eta_{1}+\vartheta_{1}\mathbf{m}\times\right)(\partial_\alpha T/T)\mathcal{D}_\alpha\mathbf{m},
\end{array}\label{ohmD-1-1}
\end{equation}
where $\mathfrak{s}$ is the reduced spin density and the form of
torques in the right hand side is dictated by the Onsager reciprocity
principle. Note that in the simplest approximation $\mathbf{H}_\mathrm{eff}$ can be calculated from an appropriate functional expressed in terms of the direction of spin density $\mathbf{m}(\mathbf{r},t)$. In a more general setting,
one may need to expand $\mathbf{H}_\mathrm{eff}$ in terms
of small $\partial_{i}T$ and $\partial_{i}\mu$.

\section{Torques by magnons from the stochastic LLG equation} 

The phenomenological
Eq.~(\ref{ohmD-1-1}) can be derived from the LLG equation with DMI. To
this end, we consider a ferromagnet with homogeneous magnetization
well below the Curie temperature. We employ the stochastic LLG equation: 
\begin{equation}
s(1+\alpha\mathbf{m}\times)\mathbf{\dot{m}}+\mathbf{m}\times(\mathbf{H}_\mathrm{eff}+\mathbf{h})=0\:.\label{eq:LLG}
\end{equation}
Here by $s$ we denote the saturation spin density and by $\mathbf{m}(\mathbf{r},t)$
a unit vector in the direction of the spin density, $\mathbf{h}$ is the random Langevin field. According to the LLG phenomenology, the effective magnetic field can be found from the Free energy, i.e.  $\mathbf{H}_\mathrm{eff}=-\delta_{\mathbf{m}}F$. In the discussion of thermal magnons,
we disregard magnetostatic and magnetocrystalline
anisotropies assuming sufficiently high temperatures, and the large $k$-limit. At very low temperatures, one needs to account for anisotropies as they lead to mixing of circular components of spin waves, and renormalization of phenomenological parameters in our theory. We consider the free energy density $F=(A/2)(\partial_\alpha \mathbf{m})^{2}+D\mathbf{m}\cdot([\mathbf{n}\times\boldsymbol{\partial}]\times\mathbf{m})-\mathbf{m}\cdot\mathbf{H}$
where $A_{ex}=A/M_{s}$ is the exchange stiffness, $D_{dm}$
describes DMI with $D\equiv D_{dm}M_{s}$, $M_{s}$ is the
saturation magnetization, and $\mathbf{H}_{e}$ is the external magnetic
field with $\mathbf{H}\equiv\mathbf{H}_{e}M_{s}$. This form of DMI
can be derived for systems with the axial symmetry around the $\mathbf{n}$-axis
and an interfacial inversion asymmetry along the $\mathbf{n}$-axis
\cite{Crepieux.Lacroix:JoMaMM1998}.

For simplicity, we assume that the slow-dynamics magnetization is
static as we can account for the time dependent effects by involving
the Onsager reciprocity principle. Without loss of generality,
we also assume a uniform temperature gradient along the $x-$axis. The vectors for the fast $\mathbf{m}_{f}(\mathbf{r},t)$
and slow $\mathbf{m}_{s}(\mathbf{r},t)$ magnetization dynamics are
related by $\mathbf{m}=(1-\mathbf{m}_{f}^{2})^{1/2}\mathbf{m}_{s}+\mathbf{m}_{f}$
where $\mathbf{m}_{s}\cdot\mathbf{m}_{f}=0$. The magnons are considered
in a coordinate system in which the $z^{'}-$axis points along the spin
density of the slow dynamics (see the figure). In this coordinate system, small excitations
will only have $m_{x}^{'}$ and $m_{y}^{'}$ components. We obtain the equation describing
the fast magnetization dynamics by linearizing the LLG equation: 
\begin{equation}
is\partial_{t}(1-i\alpha)m_{+}=\left[A(i\boldsymbol{\partial})^{2}-2iD\boldsymbol{\partial}\cdot(\mathbf{n}\times\mathbf{m}_{s})+H\right]m_{+}.\label{eq:Magnons}
\end{equation}
In the absence of anisotropy terms we can disregard the coupling between the circular components $m_{\pm}=m_{x}^{'}(\mathbf{r},t)\pm im_{y}^{'}(\mathbf{r},t)$ where $m_{x(y)}^{'}(\mathbf{r},t)$ are the transverse components  
of spin wave  \cite{Dugaev:jul2005}. As it was suggested in previous studies \cite{Moon.Seo.ea:PRB2013}, the presence of DMI in Eq.~(\ref{eq:Magnons}) leads to thermal magnons
with shifted spectrum $\omega_{k}=[H+A(\mathbf{k}+\mathbf{k}_{0})^{2}-Ak_{0}^{2}]/s$
where $\mathbf{k}_{0}=(D/A)(\mathbf{n}\times\mathbf{m}_{s})$ describes
shift in the magnon momentum induced by DMI. 

In Eq.~(\ref{eq:LLG}), we introduced the random Langevin field $\mathbf{h}$ corresponding to thermal fluctuations at temperature $T$. According to the fluctuation dissipation theorem the random fields are described by the correlator
\cite{Brown:PR1963}: 
\begin{equation}
\left\langle h_{i}(\mathbf{r},t)h_{j}(\mathbf{r}',t')\right\rangle =2\alpha sk_{B}T(\mathbf{r})\delta_{ij}\delta(\mathbf{r}-\mathbf{r}')\delta(t-t').\label{eq:stoch}
\end{equation}
We treat the fast magnetization dynamics as a linear response to the fluctuating field. In addition, we consider the slow, single-domain magnetic dynamics with long characteristic time-scale,
e.g., corresponding to ferromagnetic resonance which is typically in GHz range.

We calculate the force that fast oscillations
exert on the slow magnetization dynamics $\mathbf{m}_{s}$ by employing the method developed in Ref.~\cite{Kovalev:PRB2014}. The force
due to rapid oscillations can only come from the second order terms in $\mathbf{m}_{f}(\mathbf{r},t)$. However, direct application of the expressions from Ref.~\cite{Kovalev:PRB2014} relying on the exchange contributions results in vanishing force.  In the
effective field $\mathbf{H}_\mathrm{eff}=\mathbf{H}+A\boldsymbol{\partial}^{2}\mathbf{m}-2D[\mathbf{n}\times\boldsymbol{\partial}]\times\mathbf{m}$
only the higher order terms corresponding to DMI lead to torque in the absence of magnetic textures: 
\begin{equation}
\boldsymbol{\mathcal{T}}=-\bigl\langle\mathbf{m}_{f}\times\mathbf{H}_\mathrm{eff}\bigr\rangle=2D\bigl\langle\mathbf{m}_{f}\times[(\mathbf{n}\times\boldsymbol{\partial})\times\mathbf{m}_{f}]\bigr\rangle,\label{eq:torque}
\end{equation}
 where $\bigl\langle\ldots\bigr\rangle$ stands for averaging over
the fast oscillations induced by the random Langevin field. By analogy with the transverse spin accumulation in the context of spin-orbit torques \cite{Miron:Nature2010}, we introduce an auxiliary quantity with the meaning of the transverse spin accumulation, $\boldsymbol{\mathcal{S}}=(1/A)\boldsymbol{\mathcal{T}}\times\mathbf{m}_{s}$.
By coarse-graining various contributions, we obtain the following
expression for the transverse accumulation of magnon spins originating from DMI (i.e., the exchange term leads to a vanishing contribution): 
\begin{equation}
\boldsymbol{\mathcal{S}}=\frac{2D}{A}\bigl\langle\mathbf{m}_{f}[\mathbf{n}\cdot(\mathbf{m}_{s}\cdot\boldsymbol{\partial})\mathbf{m}_{f}]\bigr\rangle.\label{eq:spin-accumulation}
\end{equation}
We recall that we consider the reference frame with $\mathbf{z}'=\mathbf{m}_{s}$, in which vectors
$\boldsymbol{\mathcal{S}}$, $\mathbf{m}_{f}$, and $\partial_{\alpha}\mathbf{m}_{f}$
are in the $x'-y'$ plane. We can simplify expressions by switching to complex
notations $a\equiv a_{x'}^{'}+ia_{y'}^{'}$ where $\mathbf{a}$ is
an arbitrary vector in the $x'-y'$ plane. In the simplified notations, we obtain the following expression for the spin accumulation
${S}=(D/A)\bigl\langle m_{+}(\mathbf{r},t)[\boldsymbol{\upsilon}\cdot\boldsymbol{\partial}]m_{-}(\mathbf{r},t)\bigr\rangle$
where $\boldsymbol{\upsilon}\equiv(n_{z'},in_{z'},-n_{x'}-in_{y'})$
and $\boldsymbol{\partial}=(\partial_{x'},\partial_{y'},\partial_{z'})$.

Since we are interested in the steady state solution, we Fourier transform 
$m_{\pm}(\mathbf{r},t)$ 
with respect to time and transverse coordinate: 
\begin{equation}
m_{\mp}(\mathbf{q},\omega,x)=\int\frac{d^{d-1}\boldsymbol{\rho}d\omega}{(2\pi)^{d}}e^{\pm i(\omega t-\mathbf{q}\boldsymbol{\rho})}m_{\mp}(\mathbf{r},t),
\end{equation}
which leads to the expression for the spin accumulation:
\begin{equation}
\mathcal{S}=\frac{D}{A}\int\frac{d^{l}\mathbf{q}d^{l}\mathbf{q}'d\omega'd\omega}{(2\pi)^{2d}}\left\langle m_{+}(\mathbf{q},\omega,x)\boldsymbol{\upsilon}\cdot\boldsymbol{\partial}m_{-}(\mathbf{q}',\omega',x)\right\rangle,\label{eq:spin-accumulation1}
\end{equation}
 where $l=d-1$, $d=2$ or $3$ depending on the dimensionality of
the magnet and $\mathcal{S}=\mathcal{S}_{x'}+i\mathcal{S}_{y'}$ describes
two components of the spin accumulation leading to the reactive and dissipative
torques. 
 As we are interested in the linear response to the random Langevin field, the LLG Eq.~(\ref{eq:LLG}) takes the form of the following equation: 
\begin{equation}
A\left[\partial_{x}^{2}+\kappa^{2}\right]m_{-}(x,\mathbf{q},\omega)=h(x,\mathbf{q},\omega),\label{eq:Helmholtz}
\end{equation}
 where the left hand side of this equation coincides with Eq.~(\ref{eq:Magnons}) and $\kappa^{2}=[(1+i\alpha)s\omega-H]/A-k_{0}^{2}-q^{2}$.
The momentum shift by $\mathbf{k}_{0}$ discussed after Eq.~(\ref{eq:Magnons})  can be removed by
a gauge transformation, thus the effect of DMI on magnons can be accounted for by renormalizing $\kappa^{2}$. We solve Eq.~(\ref{eq:Helmholtz}) by employing the Green's function for Helmholtz equation,
$G(x-x_{0})=ie^{ik|x-x_{0}|}/(2k)$. We substitute this solution in
Eq.~(\ref{eq:spin-accumulation1}) and carry through integrations
over variables $x$ and $x_{1}$ by employing the Fourier transformed correlator for the stochastic fields \cite{Hoffman.Sato.ea:PRB2013}:
\begin{equation}
\frac{\left\langle h(x,\mathbf{q},\omega)^{*}h(x_{1},\mathbf{q}',\omega')\right\rangle }{4(2\pi)^{d}\alpha sk_{B}}=T(x)\delta(x-x_{1})\delta(\mathbf{q}-\mathbf{q}')\delta(\omega-\omega').\label{eq:stoch-1}
\end{equation} 
We arrive at the expression for the magnon spin torque: 
\begin{equation}
\mathcal{T}=-i\frac{s\alpha D}{4A^{2}}\int\frac{d^{d-1}\mathbf{q}}{(2\pi)^{d}}{\displaystyle \int_{\omega_{0}}^{\infty}}d\omega\frac{[\boldsymbol{\upsilon}\cdot\boldsymbol{\partial}]\bigl(\hbar\omega\coth\frac{\hbar\omega}{2k_{B}T(\mathbf{r})}\bigr)}{\kappa^{*}[\text{Im}(\kappa)]^{2}},\label{eq:spin-accumulation2}
\end{equation}
where we had to limit the frequency integration by $\omega_{0}=(H+Aq^{2}+Ak_{0}^{2})/s$ as  within this description we are only interested in magnonic excitations with energies
above the magnonic gap. In addition, we
replaced $2k_{B}T\rightarrow\hbar\omega\coth(\hbar\omega/2k_{B}T)$
by employing the quantum fluctuation dissipation theorem in order to introduce
a high frequency cut off at $\hbar\omega\gg k_{B}T$. Note that Eq.~(\ref{eq:spin-accumulation2}) contains terms that can be identified as magnon currents obtained from the Boltzmann approach.

We can simplify Eq.~(\ref{eq:spin-accumulation2}) after replacing
integration over $\omega$ with integration over $k$ and keeping
only the first two orders in $\alpha$, arriving at the following expression:

\begin{equation}
\mathcal{T}=-\frac{\hbar D}{A}(1+i\beta)(\boldsymbol{\upsilon}\cdot\mathbf{j}).\label{eq:Boltzmann}
\end{equation}
Here $\mathbf{j}=\mathcal{J}_{1}\boldsymbol{\partial}T/T$ coincides
with the expression for the magnon current calculated within the relaxation
time approximation where $\mathcal{J}_{n}=-\int d^{d}\mathbf{k}/(2\pi)^{d}\tau(\varepsilon)\varepsilon^{n}\upsilon_{x}^{2}\partial f_{0}/\partial\varepsilon$ and
$\tau(\varepsilon)=(2\alpha\omega)^{-1}$ account for the non-equilibrium distribution
correction $\delta f=\tau\varepsilon(\partial f/\partial\varepsilon)\upsilon_{\alpha}(\partial_{\alpha}T/T)$ arising in the Boltzmann equation
\cite{Ashcroft.Mermin:1976}. Here we also use the spectrum of magnons $\varepsilon(\mathbf{k})=\hbar(Ak^{2}+Ak_{0}^{2}+H)/s$, the velocity
$\upsilon_{x}=\partial\omega_{k}/\partial k_{x}$, and the Bose-Einstein equilibrium distribution $f_{0}=\left\{ \exp\left[\varepsilon/k_{B}T\right]-1\right\} ^{-1}$. Two terms in
Eq.~(\ref{eq:Boltzmann}) result in two torque components that are perpendicular to each other. Thus, the first term corresponds to the reactive torque and the second term corresponds to the dissipative torque, with the ratio between them given by the parameter $\beta$. As these corrections arise due to magnon spin dephasing, we expect that the parameter $\beta$ should coincides with the one obtained in the context of magnonic torques in textured ferromagnets \cite{Kovalev:PRB2014}. We can also confirm this by inspecting Eq.~(\ref{eq:spin-accumulation2}) which results in expression $\beta/\alpha=(d/2)B(x)/F_{1}(x)\sim d/2$ where $F_{n}(x)=\int_{0}^{\infty}d\epsilon\epsilon^{d/2}(\epsilon+x)^{n-1}e^{\epsilon+x}/(e^{\epsilon+x}-1)^{2}$
and $B(x)=\int_{0}^{\infty}d\epsilon(\epsilon+x)\epsilon^{d/2-1}e^{\epsilon+x}/(e^{\epsilon+x}-1)^{2}$
evaluated at the magnon gap $x=\hbar\omega_{0}/k_{B}T$ where $d=2$
or $3$. We also recall that the magnon current density is given by \cite{Kovalev:EPL2012}:
\begin{equation}
j_{\alpha}=k_{B}\partial_{\alpha}TF_{1}/(6\pi^{2}\lambda\hbar\alpha),\label{eq:MagnonCurrent-1}
\end{equation}
 where $d=3$ and $\lambda=\sqrt{\hbar A/(sk_{B}T)}$ is the thermal
magnon wavelength. In case of $d=2$ we obtain $j_{\alpha}=k_{B}\partial_{\alpha}TF_{1}/(4\pi\hbar\alpha)$.

We express the result in Eq.~(\ref{eq:Boltzmann}) in the form of the 
LLG equation which constitutes the main result of this section: 
\begin{equation}
\mathfrak{s}(1+\alpha^{s}\mathbf{m}_{s}\times)\mathbf{\dot{m}}_{s}+\mathbf{m}_{s}\times\mathbf{H}_\mathrm{eff}^{s}=\left(\eta+\vartheta\mathbf{m}_{s}\times\right)j_\alpha\mathcal{D}_\alpha\mathbf{m}_s,\label{eq:LLG-1}
\end{equation}
where $\eta=\hbar$, $\vartheta=\eta\beta$ \cite{Note}, $\mathfrak{s}=\left|\mathbf{\left\langle m\right\rangle }\right|s$
is the renormalized spin density, $\mathbf{H}_\mathrm{eff}^{s}=-(\mathfrak{s}/s)\bigl\langle\delta_{\mathbf{m}}F\bigr\rangle$
is the effective field, $\alpha^{s}=(\mathfrak{s}/s)\alpha$ is the
renormalized Gilbert damping (in other sections we always skip
the index $s$). 

\section{Critical current instability and switching} 

The LLG Eq.~(\ref{eq:LLG-1})
describes the magnon current-induced magnetic instabilities and magnetization
switching. We can estimate the corresponding critical current and
required temperature gradient by employing a simple stability analysis
of the linearized LLG equation after transforming it to Landau-Lifshitz
form, i.e., by multiplying Eq.~(\ref{eq:LLG-1}) by $(1-\alpha\mathbf{m}\times)$.
We consider the case in the figure where a time-independent
magnon current is $\mathbf{j}=j\mathbf{x}$, and the effective field
is given as $\mathbf{H}_\mathrm{eff}^{s}=H\mathbf{y}+Km_{z}\mathbf{z}$
where $H$ is the strength of the external magnetic field and $K$
is the easy-plane magnetic anisotropy, e.g., corresponding to the shape
anisotropy. When the temperature is uniform ($j=0$) at equilibrium,
the fixed point solution is $\mathbf{m}_{s}=-\mathbf{y}$. This solution becomes unstable when $j$
reaches: 
\begin{equation}
j_{c}=\frac{A}{D}\frac{H+K/2}{\left|\vartheta/\alpha-\eta\right|}.\label{eq:critcurr}
\end{equation}
We assume that $H^{2}+HK\geqslant4K^{2}\alpha^{2}$ and $\vartheta>\eta\alpha$
(for $\vartheta<\eta\alpha$ Eq.~(\ref{eq:critcurr}) gives only the
upper bound for the critical current). For $\vartheta\sim\eta\beta$
one can see that our estimate of the critical current becomes large.
When $\vartheta\sim\eta$ and $\alpha\ll1$, we obtain more favorable
estimate $j_{c}\approx\left(H+K/2\right)\alpha A/\hbar D$ which will
be used for our numerical estimates. The reason is that for a general
form of DMI corresponding to pyrochlore crystals \cite{Onose:2010}
or chiral magnets \cite{Seki:Science2012} we expect other mechanisms,
not necessarily relying on the dissipative $\beta$-type correction
\cite{Kovalev:PRB2014}, to contribute to the dissipative torque \cite{Note} by
analogy with the current-induced torques in bilayers \cite{Tserkovnyak.Bender:PRB2014}.
For estimates, we consider $\mathrm{Cu}_{2}\mathrm{OSeO}_{3}$ thin insulating
layer in ferromagnetic phase. By taking the lattice spacing $a=0.5\mbox{nm}$,
$T=50\mbox{K}$, $\alpha=0.01$, $\mathfrak{s}=0.5\hbar/a^{3}$, $A/(a^{2}k_{B})=50\mbox{K}$,
$K=4\pi M_{s}$ and $D/(ak_{B})=3\mbox{K}$ we obtain $A/D\approx80\mbox{nm}$
\cite{Seki:Science2012}. A temperature gradient comparable to experimentally
accessible $\partial_{\alpha}T=1\mbox{K}/\mu\mbox{m}$ \cite{Brandl.Grundler:APL2014}
should be sufficient for the instability according to Eq.~(\ref{eq:MagnonCurrent-1}).

\section{Magnon pumping by magnetization dynamics} The dissipative
terms in Eq.~(\ref{ohmD-1}) lead to pumping of magnons by magnetization
dynamics. This effect is analogous to charge pumping by a combination
of the spin pumping and the inverse spin Hall effect in bilayers \cite{Mosendz.Vlaminck.ea:PRB2010}
where the magnetization dynamics is induced by external
microwave fields. Under a simple circular precession of the magnetization
the time averaged magnon current is given by $j_{pump}=\sin^{2}\theta\omega\vartheta/\varrho$
where $\theta$ is the cone angle of the magnetization dynamics. We
can also define the corresponding spin current as $j_{s}=\hbar j_{pump}$.
For numerical estimates we use parameters characteristic to $\mbox{Pt}/\mbox{Co}/\mbox{AlO}_{x}$
or $\mbox{Pt}/\mbox{CoFe}/\mbox{MgO}$ thin films which we treat as
two dimensional magnets \cite{Miron.Moore.ea:NM2011}. We use $A=1.6\times10^{-11}\mbox{J}/\mbox{m}$,
$M_{s}=8.3\times10^{5}\mbox{A}/\mbox{m}$, $t=0.6\times10^{-9}\mbox{nm}$
and $D=4\times10^{-3}\mbox{J}/\mbox{m}^{2}$ arriving at $j_{s}=\sin^{2}\theta\hbar\omega D\vartheta/(4\pi At\eta\alpha)$
which is reminiscent of the expression for spin pumping $j_{s}=\sin^{2}\theta\hbar\omega g_{\uparrow\downarrow}/(4\pi)$
with $g_{\uparrow\downarrow}=D\vartheta/(At\eta\alpha)$. Compared
to spin pumping we recover an order of magnitude smaller spin current
under equal pumping conditions. 

We suppose the power dissipated by magnetization dynamics, $P=\sin^{2}\theta\omega^{2}\alpha\mathfrak{s}V$,
is equally divided between the cooled and heated reservoirs \cite{Kovalev:SSC2010}.
From Eq.~(\ref{ohmD-1}) the maximum cooling then corresponds to the
regime in which the heat current carried by magnons from the cooled
reservoir is exactly compensated by dissipation:
\begin{equation}
\frac{\triangle T}{TL}=\frac{\omega\Pi\vartheta/\varrho-\omega^{2}\alpha\mathfrak{s}L/2}{\Pi^{2}/\varrho+\kappa T}\sin^{2}\theta,\label{eq:MaxTemp}
\end{equation}
where $\kappa$ could also include the thermal conductivity of phonons, $L$ is the length of the magnet, and we assume that heat flows
couple to magnetization only via magnon currents. The maximum is reached
for $\omega=\Pi\vartheta/(\varrho\alpha\mathfrak{s}L)$. By analogy
with thermoelectric figure of merit we can define the figure of merit
for magnonic cooling as $ZT=2\triangle T_{max}/T$ which leads
to expressions: 
\begin{equation}
\frac{ZT}{\sin^{2}\theta}=\frac{(\Pi\vartheta/\varrho)^{2}/(\alpha\mathfrak{s})}{\Pi^{2}/\varrho+\kappa T}=\frac{D^{2}(\vartheta/\eta)^{2}\hbar}{6\pi^{2}\mathfrak{s}A^{2}\lambda\alpha^{2}}\frac{F_{1}^{2}}{F_{2}},
\end{equation}
where we assume sufficiently low temperature so that the effect of
phonons can be disregarded, e.g., at $3$K the thermal conductivity of magnons can become comparable to the thermal conductivity of phonons \cite{Douglass:PR1963}, $\Pi/\varrho=-\mathcal{J}_{1}$
and $\Pi^{2}/\varrho+\kappa T=\mathcal{J}_{2}$ within the relaxation
time approximation applied to thermal magnons, and $F_{1}^{2}/F_{2}\sim1$
{[}for $d=2$ we obtain $Z=D^{2}(\vartheta/\eta)^{2}\hbar F_{1}^{2}/(4\pi\mathfrak{s}A^{2}\alpha^{2}F_{2})$
where $\mathfrak{s}$ is the spin surface density{]}. For $\mbox{Pt}/\mbox{Co}/\mbox{AlO}_{x}$
or $\mbox{Pt}/\mbox{CoFe}/\mbox{MgO}$ thin films we obtain $ZT\sim0.001$.
The absolute cooling of the cold reservoir is relatively weak which
is a consequence of large dissipated power $P$. However, the relative
temperature difference between reservoirs found without $P$ can be
large at typical ferromagnetic resonance frequencies, i.e., $\triangle T/T\sim0.05$,
and it should be measurable. 

\section{Generalizations to arbitrary DMI}
The magnon spin torque in Eq.~(\ref{eq:LLG-1}) can be obtained from results in Ref.~\cite{Kovalev:PRB2014} (see Eq.~(13)) by replacing the texture derivative $\partial_i$ with a chiral derivative $\mathcal{D}_\alpha=\partial_\alpha+(D/A)(\boldsymbol{n}\times\boldsymbol{e}_\alpha)\times$\cite{Tserkovnyak.Bender:PRB2014,Kim.Lee.ea:PRL2013}. In general, such procedure does not guarantee the correct values for $\vartheta$ and $\eta$. 
Based on the phenomenological symmetry-based argument, we can describe the magnon spin torque and magnon pumping in Eqs.~(\ref{ohmD-1}) and (\ref{ohmD-1-1})  for the most general form of DMI by substituting the chiral derivative, $\mathcal{D}_\alpha=\partial_\alpha+(\bold{D}_{\alpha}/A) \times$, i.e., for the reactive and dissipative torques we obtain:
\begin{equation}
\boldsymbol{\mathcal{T}}=(1/A)\left(\eta+\vartheta\mathbf{m}\times\right) j_\alpha \bold{D}_{\alpha}\times\mathbf{m}.\label{eq:TRQ}
\end{equation}  
Here we sum over repeated indices, and the tensor $D_{\alpha \beta}=\mathbf{D}_\alpha \cdot \mathbf{e}_\beta$ describes the most general form of DMI, $F_D=D_{\alpha\beta}\varepsilon_{\beta \delta \gamma}m_\delta \partial_\alpha m_\gamma$, leading to the exchange contribution in the Free energy, $F_{ex}=(A/2)(\mathcal{D}_\alpha\mathbf{m})^{2}$. By separating $D_{\alpha \beta}$ into symmetric and antisymmetric parts, $D_{\alpha \beta}=D_{\alpha \beta}^{sym}+\varepsilon_{\alpha \beta \gamma} D_\gamma^{ant}$, we identify $\mathbf{D}^{ant}=D\mathbf{n}$ for DMI due to structural asymmetry in the figure. The contribution $D_{\alpha \beta}^{sym}=D \delta_{\alpha \beta}$ arises in non-centrosymmetric crystals, e.g.,  in $\mathrm{Cu}_{2}\text{OSeO}_{3}$, resulting in $\boldsymbol{\mathcal{T}}=(D/A)\left(\eta+\vartheta\mathbf{m}\times\right) \mathbf{j}\times\mathbf{m}$.

\section{Conclusions}

We developed a phenomenological description for
the interplay between magnetization dynamics and magnon currents in ferromagnets
with DMI. Our theory describes: (i)
magnon current-induced instability and switching; (ii) pure spin current
pumping; and (iii) cooling effects in single-domain magnets. The strength
of all mentioned effects is related to the magnitude of dissipative
torque which is weakened by the Gilbert damping factor compared to
the reactive torque for the simple form of DMI considered here. The
general form of DMI, $\mathbf{D}_{ij}\cdot(\mathbf{S}_{i}\times\mathbf{S}_{j})$,
should in principle result in a situation where the dissipative and
reactive torques are comparable \cite{Note} in analogy to spin-orbit torques in metal/ferromagnet bilayers.
We thus expect magnonic dissipative torques and spin pumping in
such systems as pyrochlore crystals (e.g., $\mathrm{Lu}_2 \mathrm{V}_2 \mathrm{O}_7$) and chiral magnets (e.g., $\mathrm{MnSi}$ or $\mathrm{Cu}_{2}\text{OSeO}_{3}$). The proposed
pumping mechanism could potentially be useful for electronics relying
on pure spin currents. Our results agree with Ref.~\cite{Manchon.Ndiaye.ea:2014} where only the reactive thermal torque has been discussed in detail.

\acknowledgments
We are grateful to K. Belashchenko and C. Binek for discussions. This work was supported in part by the NSF under Grants No. Phy-1415600, No. DMR-1420645,  and NSF-EPSCoR 1004094, and performed in part at the Central Facilities of the Nebraska
Center for Materials and Nanoscience, supported by the Nebraska Research Initiative.

\bibliographystyle{eplbib}
\bibliography{MyBIB-C,Berryheat}

\begin{thebibliography}{40}
\bibitem{Note}
In our discussion the strength of the dissipative torque is proportional to the
  dephasing parameter $\beta$ while it does not have to be so for a general
  magnonic band with non-trivial Berry curvature.
\end{thebibliography}

\end{document}